\documentclass[a4paper,10pt]{amsart}
\usepackage{amsmath}
\usepackage{amsthm}
\usepackage{graphicx}

\newtheorem{statement}{Statement}
\title{Quasisolution of the inverse 3D aerohydrodynamics problem}
\author{Pyotr N. Ivanshin}

\begin{document}

\maketitle

\begin{abstract}
In the article we generalise the quasisolution approach to the planar aerohydrodynamics problems to 3D case.
We search for solution in the form of the linear spline.
\end{abstract}

\section{Introduction}
In the article we generalise the quasisolution approach to the planar aerohydrodynamics problems to 3D case.
The first step in solution of the planar problems was made by Mangler \cite{Mang}. Later Tumashev rediscovered this approach \cite{Tum}. The main problem in the inverse solution construction is that there exist certain conditions of both mathematical and mechanical origin, i.e. the reconstructed contour must be a closed Jordan curve and the velocity at infinity must equal some specified value. In order to overcome these difficulties researchers introduced numerous quasisolutions \cite{AM}, \cite{PI}.

Here we try to extend the plaar constructions to 3D case. We reduce the 3D problem to the set of planar ones similarly to the cases of the other 3D problems \cite{ell}, \cite{book}. So here we have something close but not identical to blade-to-blade analysis widely applied to the problem under consideration. The blades under consideration are not mutually independent since we consider the flow component transversal to the blades. 
We search for solution in the form of the spline linear on the transversal to blade coordinate.

\section{Main Results}

Let us consider the set of equations:
$$
\frac{\partial u}{\partial x}+\frac{\partial v}{\partial y}+\frac{\partial w}{\partial h}=0,
$$
$$
\frac{\partial u}{\partial y}=\frac{\partial v}{\partial x}, \, \frac{\partial u}{\partial h}=\frac{\partial w}{\partial x}, \, \frac{\partial v}{\partial h}=\frac{\partial w}{\partial y}
$$
Let us search for the solution in the form of the linear on $h$ polynomial
\begin{eqnarray*}
u(x, y, h)=u_0(x, y)+h u_1 (x, y),\\
v(x, y, h)=v_0(x, y)+h v_1 (x, y),\\
w(x, y, h)=w_0(x, y)+h w_1 (x, y).
\end{eqnarray*}
Then we have the following relations on $u_j, v_j, w_j$, $j=0, 1$:
\begin{eqnarray}\label{uv1}
\frac{\partial u_1}{\partial x}=-\frac{\partial v_1}{\partial y},
\frac{\partial u_1}{\partial y}=\frac{\partial v_1}{\partial x}.
\end{eqnarray}
\begin{eqnarray}\label{uv0}
\frac{\partial u_0}{\partial x}=-\frac{\partial v_0}{\partial y}-w_1,
\frac{\partial u_0}{\partial y}=\frac{\partial v_0}{\partial x}.
\end{eqnarray}
\begin{eqnarray}
\frac{\partial w_0}{\partial x}+h \frac{\partial w_1}{\partial x}=u_1,
\frac{\partial w_0}{\partial y}+h \frac{\partial w_1}{\partial y}=v_1.
\end{eqnarray}
The last set of relations implies that $w_1=\mathrm{const}$. Thus we obtain
\begin{eqnarray}\label{w0}
\frac{\partial w_0}{\partial x}=u_1,
\frac{\partial w_0}{\partial y}=v_1.
\end{eqnarray}

System of equations (\ref{w0}) is correct due to the second equation of system (\ref{uv1}). Let us introduce the usual complex variable $z=x+i y$. Thus we obtain the solution 
$$
 w_0=\frac{1}{2 i}(\int v_1(z)+ i u_1 (z) d z -\overline{\int v_1(z)+ i u_1 (z) d z}).
$$
So this function is up to some real constant defined by the velocities $u_1$ and $v_1$.

This linear spline allows us to obtain two adjacent airfoil profiles. So we have the unknown functions $u_0$, $v_0$, $u_1$, $v_1$ and $w_0$ and one unknown constant $w_1$. Assume that we fix the velocity values at these adjacent blades. Then for the functions $u_1$ and $v_1$ we have the usual plane problem. This allows us to reconstruct the analytic function $f_1(z)=v_1(z)+i u_1(z)$. The analytic function constructed by $u_0$ and $v_0$ we can find up to the summand depending on $w_1$:  $f_0(z)=v_0(x, y)+i u_0(x, y)+i w_1(x-i y)$. 
% Then the potential is a $2$-harmonic function. 

It seems natural for the gas or fluid particles to travel along geodesic lines on the surface of the airfoil. Because of this  we necessarily obtain the additional condition: the geodesic lines on the upper and lower surfaces of the airfoil must have the same start and end points. Also we naturally must have 
\begin{equation} \label{branch}
w(x, y, h)|_{B}\cong 0.
\end{equation}
Here $B$ is flow branch point.

We now have a system of $4$ real-valued equations on the analytic functions  $h f_1+f_0$ and $f_0$. These problems are standart inverse aero-hydrodynamics problems. We note
that modified system \ref{uv0} is closely related to classical system \ref{uv1}. Even the functions $f_0$ and $f_1$ are both harmonic. The only difference in the solution is that the velocity potential $\omega_0$ for the function $f_0$  equals $\phi_0(z)+i \psi_0(z)+i w_1 |z|^2$.  So for the cut we must have the following initial data: the tangent to the cut velocity $V$ depending on the cut length coordinate, the transversal to the cut velocity $w_1$ and the  velocities $V$ and $w_1|s|$ sum. The last notion is exactly what we apply to solve the classical problems.  So we first obtain $f_0$ and $f_1$ then deduce from them $i w_1 \overline{z}$ and find $u_0$, $v_0$, $u_1$ and $v_1$. The latter velocity components allow us to reconstuct the contours.

\begin{figure}[h] \label{fig:1}
 \centering
 \includegraphics[height=40mm]{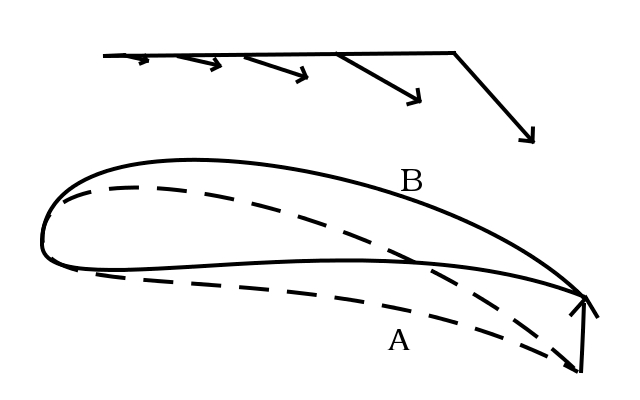}
 % imagelong.png: 500x300 pixel, 72dpi, 17.64x10.58 cm, bb=0 0 500 300 
 \caption{A --- Contour reconstructed from equations with the help of the analytic function, B --- Contour reconstructed under $w_1$ shift}

\end{figure}

% since the summand $i w_1 |z|^2$ is purely imaginary and does not affect $\phi_0(z)$. 

The solutions of the system depend on parameter $w_1$.  The simplest and most natural way to solve the equations is to determine the value of $w_1$ from the initial conditions.

Then we construct $w_0$ with the help of condition (\ref{branch}) and boundary conditions. 

The next step is to attach the adjacent section to the first one. In general the linear spline does not allow us to glue togehter values of $u$ and $v$ on the common blade so we simply ignore this blade and solve the problem only for the newly added third blade.

% So there are two solution methods. The first is to obtain $w_1$ fromdata. The second is to compute a set of parameter-dependent solutions and choose the optimal one. Here we find $w_1$ from condition (\ref{branch}). 

Let us now turn to the problem of mutual positioning of the adjacent blades. Note first that in general even values of the velocity coordinates may not provide us with the directions of the planes tangent to the constructed surface. For example, in the case of $w_0=w_1=0$ we have absolutely no information on this matter. Also these planes may not define an integable distribution  (see Frobenius theorem). Thus we need to introduce another way of this positioning.

The easiest mathod is the well-known least square method. We apply this method to the set of lengths between points on the contours. Since we obtain the contours as the set of points $C_1: \{(x_1^1, y_1^1), \ldots, (x_1^n, y^n_1)\}$, $C_2: \{(x_2^1, y_2^1), \ldots, (x_2^n, y^n_2)\}$ we consider the function $f(x, y)=\sum\limits_{i=1}^{n} ((x_i^1-x_i^2+x)^2+(y_i^1-y_i^2+y)^2)$. Then we find the minimal point $(x_0, y_0)$ of this function. Thus we minimize the mean distance between the contours.

\begin{statement}
If the contours $C_1$ and $C_2$ are similar then the limit optimal point also minimizes the square of the ruled surface $S$ between $C_1$ and $C_2$. 
\end{statement}

\begin{proof}
Let us consider the triangulation of the surface $S$ so that the triangles has either vertices $(x_1^i, y^i_1)$, $(x_1^{i+1}, y^{i+1}_1)$, $(x_2^i, y^2_1)$ or  $(x_2^i, y^i_2)$, $(x_2^{i+1}, y^{i+1}_2)$, $(x_1^i, y^1_1)$. Let us denote the first triangle as $\Delta_1^i$ and the second as $\Delta_2^i$.

Then the surface is the sum of the sums $S_1=\sum\limits_{i=1}^{n}S(\Delta_1^i)$ and  $S_2=\sum\limits_{i=1}^{n}S(\Delta_2^i)$ limits.

Note that since the contours are similar we have $S(\Delta_2^i)=\alpha S(\Delta_1^i)$ for any $i=1, \ldots, n$.

The first sum as $n \to \infty$ turns into the integral 
$$
\int\limits_{S_1} \sqrt{(x^1(t)-x^2(t)+x)^2+(y^1(t)-y^2(t)+y)^2} d t
$$
The minimum of this integral happens at the same point as the minimum of the function $f(x, y)$.

\end{proof}

Example 1.

Let us consider the standart velocity distribution.

We obtain the following contours:

\begin{figure}[h] \label{fig:1}
 \centering
 \includegraphics[height=40mm]{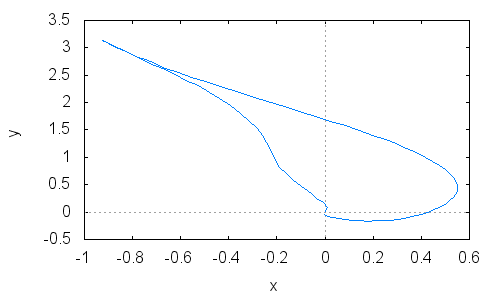}
 % imagelong.png: 500x300 pixel, 72dpi, 17.64x10.58 cm, bb=0 0 500 300 
 \caption{The lower (longer) section}

\end{figure}

\begin{figure}[h]\label{fig:2}
 \centering
 \includegraphics[height=40mm]{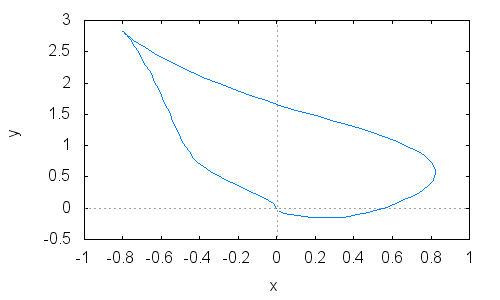}
 % imageshort.png: 500x300 pixel, 72dpi, 17.64x10.58 cm, bb=0 0 500 300
 \caption{The upper (shorter) section}
\end{figure}

Their mutual shift is given by numbers $a=-0.089794872821941$, being the shift along $OX$ axis, and $b=0.191236618879538$, being the shift along $OY$ axis. This means that the point with coordinates $(0, 0)$ on the second graph has coordinates $(a, b)$ in the coordinate system of the first graph.

One other possible way to find mutual blades placement is to maximize the lift force. In order to do this one must maximize the function $f(x, y)=\sum\limits_{i=1}^{k} ((x_i^1-x_i^2+x)^2+(y_i^1-y_i^2+y)^2)^{1/2} (v_i^1+v_i^2)-\sum\limits_{i=k+1}^{n} ((x_i^1-x_i^2+x)^2+(y_i^1-y_i^2+y)^2)^{1/2} (v_i^1+v_i^2)$. Here the nodes $(x_i^1, y_i^1)$, $(x_i^2, y_i^2)$, $i=1, \ldots, k$ lie on the lower part of the wing and the nodes $(x_i^1, y_i^1)$, $(x_i^2, y_i^2)$, $i=k+1, \ldots, n$ belong to the upper part of the wing. Since the velocity on the upper wing part is generally greater than that on the lower the wing end must be sloped lowerwise. 

\section{Spline of degree $2$}

Let us search for the solution in the form of the $h$ polynomial quadratic on $h$:
\begin{eqnarray*}
u(x, y, h)=u_0(x, y)+h u_1 (x, y)+h^2 u_2 (x, y),\\
v(x, y, h)=v_0(x, y)+h v_1 (x, y)+h^2 v_2 (x, y),\\
w(x, y, h)=w_0(x, y)+h w_1 (x, y)+h^2 w_2 (x, y).
\end{eqnarray*}
Then similarly to the linear case we have the following relations on $u_j, v_j, w_j$, $j=0, 2$:
\begin{eqnarray}\label{uv2}
\frac{\partial u_2}{\partial x}=-\frac{\partial v_2}{\partial y},
\frac{\partial u_2}{\partial y}=\frac{\partial v_2}{\partial x}.
\end{eqnarray}
\begin{eqnarray}\label{uv12}
\frac{\partial u_1}{\partial x}=-\frac{\partial v_1}{\partial y}-2 w_2,
\frac{\partial u_1}{\partial y}=\frac{\partial v_1}{\partial x}.
\end{eqnarray}
\begin{eqnarray}\label{uv02}
\frac{\partial u_0}{\partial x}=-\frac{\partial v_0}{\partial y}- w_1,
\frac{\partial u_0}{\partial y}=\frac{\partial v_0}{\partial x}.
\end{eqnarray}
Hence the eldest coefficient
\begin{equation}
w_2=\mathrm{const}, 
\end{equation}
\begin{eqnarray}
\frac{\partial w_1}{\partial x}=2 u_2,
\frac{\partial w_1}{\partial y}=2 v_2,
\end{eqnarray}
\begin{eqnarray}
\frac{\partial w_0}{\partial x}=u_1,
\frac{\partial w_0}{\partial y}=v_1.
\end{eqnarray}
Also we have harmonic functions $w_1(z, \overline{z})=\frac{1}{2 i}(\int(v_2+i u_2) d z-\overline{\int(v_2+i u_2) d z}$, $w_0(z, \overline{z})=\frac{1}{4 i}(\int(v_1+i u_1) d z-\overline{\int(v_1+i u_1) d z}$.

Assume that we construct our solution only for two blades. Then we have only two boundary conditions on three anaytic functions $f_0$, $f_1$ and $f_2$. Note that $w_0$ is again defined by $u_0$ and $v_0$ and does not take part in the boundary data. This leaves us with the function $w_1$ which allows us to introduce one real constant and $w_2$ which itself is a real constant. We choose $w_1$ for the second section to be equal to $w_1+w_2$ of the first section. We find the value of $w_2$ for the second section from the initial data.  So this construction allows us to glue the velocity components $u$ and $v$ values on the common blade of the adjacent sections.

\end{document}